\documentclass[aip,reprint]{revtex4-2}

\usepackage{amsmath,amsfonts,amssymb,amsthm}
\usepackage{graphicx}
\usepackage{braket}
\usepackage{hyperref}

\renewcommand{\vec}[1]{\boldsymbol{#1}}
\newcommand{\mtx}[1]{\boldsymbol{\mathsf{#1}}}

\newif\ifshowfigures
\showfigurestrue 

\begin{document}

\title{Classical model of quantum interferometry tests of macrorealism}

\author{Brian R. La Cour}
  \email{blacour@arlut.utexas.edu}
  \affiliation{Applied Research Laboratories, 
               The University of Texas at Austin, 
               P.O. Box 8029, 
               Austin, TX 78713-8029, USA}
               
\date{\today} 

\begin{abstract}
Macrorealism is a characteristic feature of many, but not all, classical systems.  It is known, for example, that classical light can violate a Leggett-Garg inequality and, hence, reject a macrorealist interpretation.  A recent experiment has used entangled light and negative measurements to demonstrate a loophole-free test of macrorealism [\textit{PRX Quantum} \textbf{3}, 010307 (2022)].  This paper shows that such an experiment, while soundly rejecting macrorealism, may nevertheless be open to a classical interpretation.  This is done by offering an explicit classical model of heralded photon detection in an optical interferometer with beam blockers.  A numerical analysis of the model shows good agreement with experimental observations and consistency with both local realism and a rejection of macrorealism.
\end{abstract}

\maketitle


\section{Introduction}

Albert Einstein famously asked whether the Moon is still there when no one looks.\cite{Pais1979}  This, in essence, is the question of macrorealism, and in systems such as the Moon the answer seems quite obvious.   Although quantum mechanics has challenged our notion of local realism in the microscopic realm, the macroscopic world has largely seemed a refuge from such speculations.

The term \emph{macroscopic realism} (or, macrorealism) was first introduced by Anthony Leggett and Anupam Garg to describe what they saw as a key feature of the classical worldview.\cite{Leggett1985}  They defined it by two fundamental properties: \emph{Macroscopic realism per se:} ``A macroscopic system with one or more macroscopically distinct states available to it will at all times \emph{be} in one or the other of these states. [Emphasis in original]''  and \emph{Noninvasive measurability:} ``It is possible, in principle, to determine the state of the system with arbitrarily small perturbations on its subsequent dynamics.''\cite{Leggett1985}  Later, Kofler and Brukner suggested \emph{no-signaling in time} as an alternative necessary condition for defining macrorealism that is useful for testing noninvasiveness.\cite{Kofler2013, Halliwell2017}

In the context of optical interferometry, the subject of this paper, the macroscopically distinct states are taken to be the two exit ports of a beam splitter.  By considering a network of beam splitters, one may define several such macroscopically distinct states, with each split in the network constituting a different ``time'' when the macrostate might have changed.  One may then define correlations of these macrostates across different times.  An inequality derived from Leggett and Garg's work gives an upper bound for certain combinations of these correlations.\cite{Leggett1985}  Several quantum mechanical systems have been shown to violate this bound, suggesting a rejection of macrorealism for these systems.\cite{Emary2014, Robens2015, Zhou2015, Formaggio2016, Knee2016, Wang2018, Ku2020}  While potential loopholes have been identified for these experiments it is believed that loophole-free violations have recently been demonstrated using interaction-free measurements.\cite{Wilde2012, Joarder2022}

The separate notions of macrorealism and realism are sometimes conflated, and this can lead to misunderstandings.  For example, Joarder \textit{et al.} assert that macrorealism is ``a generic trait characterizing any classical description of the physical world.''\cite{Joarder2022}  On the other hand, Chevalier \textit{et al.} have argued that classical light can violate a Leggett-Garg inequality (LGI),\cite{Chevalier2021} and this has indeed been demonstrated experimentally.\cite{Zhang2018,Zhang2021}

It should not be surprising that classical light can violate an LGI, as ``classical wave mechanics is not a macroscopic-real theory.''\cite{Emary2014}  The experiments of Joarder \textit{et al}., however, go much further in demonstrating an LGI violation with entangled light, using heralded photons from a parametric downconversion source.\cite{Joarder2022}  The experiment uses a clever system of beam blockers to perform interaction-free negative measurements, a technique first described by Leggett and later used in the context of the Elitzur-Vaidman bomb experiment.\cite{Leggett1988, Elitzur1993,Kwiat1995}  While this experiment provides a convincing rejection of macrorealism, it is unclear whether a classical interpretation of the results might not be possible.

This paper attempts to answer this question by offering an explicit classical model.  The basis for the model is quite simple and has been described elsewhere.\cite{LaCour&Yudichak2021a,LaCour&Yudichak2021b}  Specifically, entangled light is modeled as multi-mode squeezed light using real, rather than virtual, vacuum modes, and photon detection is modeled as a deterministic process based on amplitude threshold crossing events.\cite{LaCour&Williamson2020}  This introduces a nonlinear element to the measurement process and is important for performing the post-selection needed to produce contextuality.\cite{Leifer2005}

The outline of the paper is as follows.  Section \ref{sec:experiment} describes the experimental setup and data analysis procedures.  Section \ref{sec:LGI} provides a review of the Leggett-Garg inequalities, emphasizing the mathematical definition of macrorealism and its relation to both the validity of the inequalities and how they are tested experimentally.  The classical model itself is described in Sec.\ \ref{sec:model}.  The model is analyzed numerically in Sec.\ \ref{sec:results}, where simulation results are compared to the experimental results of Ref.\ \onlinecite{Joarder2022}.  A discussion of the detection efficiency loophole is given in Sec.\ \ref{sec:loophole}, and conclusions are summarized in Sec.\ \ref{sec:conclusion}.


\section{Interferometry Experiment}
\label{sec:experiment}

Using a realist wave model of light immediately raises the question of how one even defines a macroscopically distinct state for the purpose of testing an LGI.  Rather than attempt to provide a model-based definition, an operational one will be used based on the experimental procedure described in Ref.\ \onlinecite{Joarder2022}.  The approach taken therefore will be to treat the model as a kind of black box for which only the measurement outcomes (i.e., detector clicks) and their associated measurement contexts will be considered in the analysis.  Note that whether one considers a wave-based or particle-based notion of light is immaterial to the execution and analysis of the experiment.  What follows is a description of the notional experiment and the analysis procedure.

\begin{figure}[ht]
\includegraphics[width=\columnwidth]{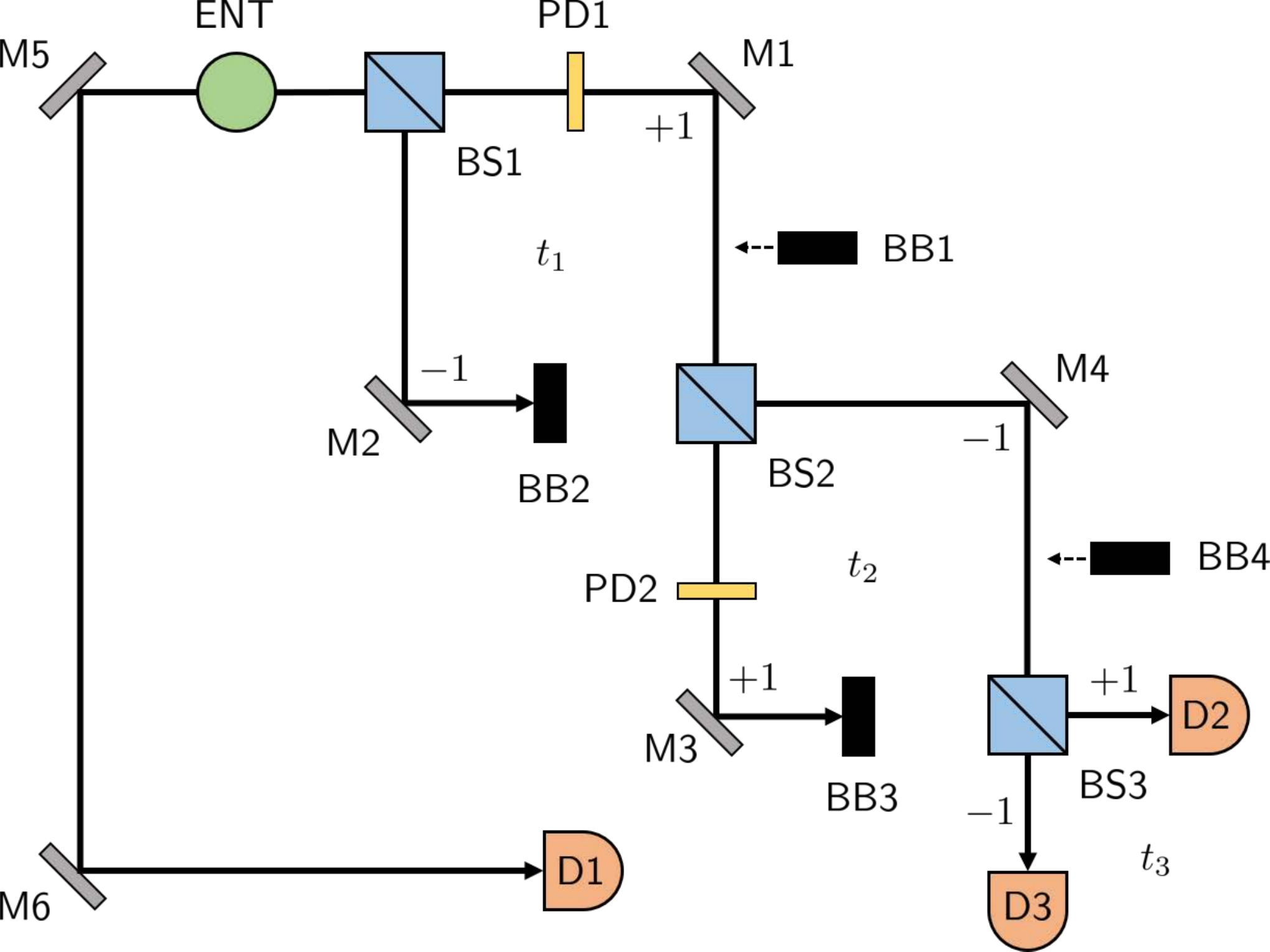}
\caption{(Color online) Experimental setup for an optical test of the Leggett-Garg inequality.}
\label{fig:setup}
\end{figure}

A simplified picture of the experimental setup is shown in Fig.\ \ref{fig:setup}.  The entanglement source (\textsf{ENT}) produces twin beams of polarization entangled light traveling left and right.  The left beam goes to detector \textsf{D1} and is used for heralding.  The right beam enters a chained pair of unbalanced Mach-Zehnder interferometers.  The beam splitters \textsf{BS1}, \textsf{BS2}, \textsf{BS3} have transmittance $T_1, T_2, T_3$, respectively, and reflectance $R_i = 1 - T_i$.  The phase delays \textsf{PD1} and \textsf{PD2} produce phase shifts of $\theta_1$ and $\theta_2$, respectively.  The paths exiting each of the three beam splitters are construed as the macroscopic states such that the upper (lower) arm of beam splitters \textsf{BS1} and \textsf{BS3} corresponds to a value of $+1$ ($-1$), while the upper (lower) arm of beam splitter \textsf{BS2} corresponds to a value of $-1$ ($+1$).  Each pair of arms exiting beam splitters \textsf{BS1}, \textsf{BS2}, \textsf{BS3} corresponds to times $t_1$, $t_2$, $t_3$, respectively.  For heralding detections at \textsf{D1}, we consider coincident detections at either \textsf{D2} or \textsf{D3}, but not both, corresponding to measurements of $+1$ or $-1$, respectively, for $t_3$.

To perform negative measurements at times $t_1$ and $t_2$, a set of four beam blockers is employed, which may be inserted in or removed from each of the four corresponding paths.  Beam blockers \textsf{BB1} and \textsf{BB2} correspond to the $+1$ and $-1$ arms, respectively, exiting \textsf{BS1}, while beam blockers \textsf{BB3} and \textsf{BB4} correspond to the $+1$ and $-1$ arms, respectively, exiting \textsf{BS2}.  An insertion of, say, beam blockers \textsf{BB2} and \textsf{BB3}, as illustrated in Fig.\ \ref{fig:setup}, would allow only the $+1$ path for $t_1$ and the $-1$ path for $t_2$.  If, in this circumstance, a heralded detection on, say, \textsf{D3} is obtained, then it is considered a realization of the measurement outcome $(q_1, q_2, q_3) = (+,-,-)$, where $\pm1$ has been abbreviated as $\pm$ for convenience.  

Note that $q_1 = +$  and $q_2 = -$ are not actually measurement outcomes but, rather, describe a particular measurement context.  They may be construed as measurements only insofar as one adopts a particle picture of light in which an individual ``photon'' exits the beam splitter in exactly one of the two directions.  This, of course, is the macrorealist view we are attempting to refute.  We may describe each possible measurement context with a binary vector $\vec{b} = (b_1, b_2, b_3, b_4)$, where $b_i = 0$ ($b_i = 1$) indicates presence (absence) of the $i^{\rm th}$ beam blocker.  The above example with \textsf{BB2} and \textsf{BB3} inserted, for example, would correspond to the measurement context $\vec{b} = (1,0,0,1)$.

The experiment proceeds by arranging a particular measurement context and measuring the coincident counts on \textsf{D1} and either \textsf{D2} (giving $q_3 = +$) or \textsf{D3} (giving $q_3 = -$).  There are three types of experiments.  The first measures $N_{t_1,t_3}(q_1,q_3)$ and uses either $\vec{b} = (1,0,1,1)$ (for $q_1 = +$) or $\vec{b} = (0,1,1,1)$ (for $q_1 = -$).  Specifically, $N_{t_1,t_3}(+,-)$, say, is the number of coincident counts on \textsf{D1} and \textsf{D3} (but not \textsf{D2}) when only \textsf{BB2} is inserted.  The second measures $N_{t_2,t_3}(q_2,q_3)$ and uses either $\vec{b} = (1,1,1,0)$ (for $q_2 = +$) or $\vec{b} = (1,1,0,1)$ (for $q_2 = -$).  The third measures $N_{t_1,t_2,t_3}(q_1,q_2,q_3)$ and inserts one of four combinations of two beam blockers as follows: (1) $\vec{b} = (1,0,1,0)$ for $(q_1,q_2) = (+,+)$, (2) $\vec{b} = (1,0,0,1)$ for $(q_1,q_2) = (+,-)$, (3) $\vec{b} = (0,1,1,0)$ for $(q_1,q_2) = (-,+)$, and (4) $\vec{b} = (0,1,0,1)$ for $(q_1,q_2) = (-,-)$.  These define all of the measurements performed.  All that remains is data analysis to construct probability distributions and compute correlations.

From $N_{t_1,t_3}(q_1,q_3)$ we may construct a probability mass function (PMF) $P_{t_1,t_3}(q_1,q_3) \propto N_{t_1,t_3}(q_1,q_3)$ such that
\begin{equation}
P_{t_1,t_3}(q_1,q_3) = \frac{N_{t_1,t_3}(q_1,q_3)}{\sum_{q'_1,q'_3}N_{t_1,t_3}(q'_1,q'_3)} \; .
\end{equation}
Note that the more formal term ``probability mass function'' is used to emphasize that, although $P_{t_1,t_3}$ satisfies the mathematical requirements of a discrete probability distribution, it does not \textit{per se} describe the frequencies of random events since $q_1$ describes a measurement context rather than a measurement outcome.  The PMFs
\begin{equation}
P_{t_2,t_3}(q_2,q_3) \propto N_{t_2,t_3}(q_2,q_3)
\end{equation}
and
\begin{equation}
P_{t_1,t_2,t_3}(q_1,q_2,q_3) \propto N_{t_1,t_2,t_3}(q_1,q_2,q_3)
\end{equation}
are defined similarly.  Finally, from $P_{t_1,t_2,t_3}(q_1,q_2,q_3)$ one can define the marginal PMF
\begin{equation}
P_{t_1,t_2}(q_1,q_2) = \sum_{q'_3} P_{t_1,t_2,t_3}(q_1,q_2,q'_3) \; .
\label{eqn:t1t2}
\end{equation}

The Leggett-Garg statistic is now computed as
\begin{equation}
K = C_{t_1,t_2} + C_{t_2,t_3} - C_{t_1,t_3} \; ,
\label{eqn:LGS}
\end{equation}
where
\begin{equation}
C_{t_i,t_j} = \sum_{\pm} P_{t_i,t_j}(\pm,\pm) - \sum_{\pm} P_{t_i,t_j}(\pm,\mp) \; .
\label{eqn:corr}
\end{equation}
A value of $K > 1$ is considered a violation of the Leggett-Garg inequality.  A related quantity,
\begin{equation}
W = P_{t_1,t_3}(-,+) - P_{t_2,t_3}(-,+) - P_{t_1,t_2}(-,+) \; ,
\label{eqn:WLGS}
\end{equation}
is used to test the Wigner form of the Leggett-Garg inequality (WLGI).\cite{Saha2015}  A value of $W > 0$ is considered a violation of the WLGI.

It may seem curious that the aforementioned PMFs refer only to one random outcome (i.e., whether $q_3 = +$ or $q_3 = -$).  One may equally adopt an interpretation in terms of a postive operator-valued measure (POVM) generated by random measurement configurations.  For example, to measure $P_{t_1,t_3}(q_1,q_3)$ we might flip a coin to determine whether to measure $q_1 = +$ or $q_1 = -$.  Having chosen the corresponding measurement configuration, we then measure $q_3$, taking care to post-select only valid coincident events between \textsf{D1} and either \textsf{D2} or \textsf{D3} (but not both).  In this way, we obtain four mutually exclusive events that, together, cover all possible outcomes.  This justifies the normalization of $N_{t_1,t_3}(q_1,q_3)$ used to obtain $P_{t_1,t_3}(q_1,q_3)$.  The other two PMFs may be interpreted similarly.


\section{The Leggett-Garg Inequality}
\label{sec:LGI}

Suppose we are given PMFs $P_{t_1,t_3}(q_1,q_3)$, $P_{t_2,t_3}(q_2,q_3)$, and $P_{t_1,t_2,t_3}(q_1,q_2,q_3)$.  From these we may define $P_{t_1,t_2}(q_1,q_2)$, as given by Eqn.\ (\ref{eqn:t1t2}).  In addition, we may define the following two marginal PMFs:
\begin{align}
\tilde{P}_{t_1,t_3}(q_1,q_3) &= \sum_{q'_2} P_{t_1,t_2,t_3}(q_1,q'_2,q_3) \; , \\
\tilde{P}_{t_2,t_3}(q_2,q_3) &= \sum_{q'_1} P_{t_1,t_2,t_3}(q'_1,q_2,q_3) \; .
\end{align}
Now, there is no reason, \textit{a priori}, to suppose that either $\tilde{P}_{t_1,t_3} = P_{t_1,t_3}$ or $\tilde{P}_{t_2,t_3} = P_{t_2,t_3}$.  This, however, is the assumption needed to prove the Leggett-Garg inequality and, hence, may be considered the mathematical expression of macrorealism.  If one adopts the notion of light as being composed of discrete, localized corpuscles (the physical description of macrorealism in this context), then it is natural to suppose equality, as the marginals are merely a sum over mutually exclusive events.  If one adopts a wave-like picture of light, however, such an assumption is clearly unwarranted.

To derive the inequality, let us consider, in addition to the correlations $C_{t_i,t_j}$ defined in Eqn.\ (\ref{eqn:corr}), the correlations of the marginal probabilities, given by
\begin{equation}
\tilde{C}_{t_i,t_j} = \sum_{\pm} \tilde{P}_{t_i,t_j}(\pm,\pm) - \sum_{\pm} \tilde{P}_{t_i,t_j}(\pm,\mp) \; .
\end{equation}
Writing the marginals in terms of $P_{t_1,t_2,t_3}$ and collecting terms, it is straightforward to show that
\begin{equation}
C_{t_1,t_2} + \tilde{C}_{t_2,t_3} - \tilde{C}_{t_1,t_3} \le 1 \; .
\label{eqn:LGI}
\end{equation}
This is the Leggett-Garg inequality (LGI).  The crux of the result can be described more succinctly by the related Wigner form of the LGI (WLGI), given by
\begin{equation}
\tilde{P}_{t_1,t_3}(-,+) - \tilde{P}_{t_2,t_3}(-,+) - P_{t_1,t_2}(-,+) \le 0 \; .
\label{eqn:WLGI}
\end{equation}

Colloquially, we say that $K > 1$ is a ``violation of the Leggett-Garg inequality,'' but this should not be misconstrued as a violation of Eqn.\ (\ref{eqn:LGI}).  Equations (\ref{eqn:LGI}) and (\ref{eqn:WLGI}) are mathematical statements that, of course, cannot be violated.  By contrast, the Leggett-Garg statistic, $K$, defined in Eqn.\ (\ref{eqn:LGS}) is not constrained by this bound.  If indeed we do find that $K > 1$, then we may validly conclude that $\tilde{P}_{t_1,t_3} \neq P_{t_1,t_3}$ or $\tilde{P}_{t_2,t_3} \neq P_{t_2,t_3}$.  It should not be surprising that a wave-based model of light should be capable of yielding this result.  Such a model, motivated by a physical description of quantum light, is discussed in the next section.


\section{Classical Model Description}
\label{sec:model}

The model considered here has been discussed elsewhere and is based on the Gaussian nature of entangled light arising from parametric downconversion.\cite{LaCour&Yudichak2021a}  For such systems, the Wigner function over the phase space variables constitutes a valid probability density function.  One may therefore adopt a mathematically equivalent description in terms of random variables, the form of which is given by the Bogoliubov transformations of the corresponding operators.  In particular, the lowering operators for each mode correspond to the complex amplitudes of the Jones vectors for each of the outgoing beams of the entanglement source.  This will now be defined explicitly, along with a description of the detector model.

Consider an entanglement source modeled as a set of complex Gaussian random variables whose joint distribution may be identified with an entangled multi-mode squeezed vacuum state.  Let $\vec{a}_1(t_0)$ denote the $2\times1$ Jones vector for the left-traveling beam at time $t_0$ and let $\vec{a}_2(t_0)$ be similarly defined for the right-traveling beam.  For a single coherence time, these are given by
\begin{subequations}
\begin{align}
\vec{a}_1(t_0) &= \sigma \vec{z}_1 \cosh r + \sigma \vec{z}_2^* \sinh r \\
\vec{a}_2(t_0) &= \sigma \vec{z}_2 \cosh r + \sigma \vec{z}_1^* \sinh r \; ,
\end{align}
\label{eqn:entanglement}
\end{subequations}
where $\sigma = 1/\sqrt{2}$ gives the scale of the vacuum fluctuations, $r \ge 0$ is the squeezing strength, and $\vec{z}_1$, $\vec{z}_2$ are independent standard complex Gaussian random vectors (i.e., $\mathsf{E}[\vec{z}_i] = \vec{0}$, $\mathsf{E}[\vec{z}_i \vec{z}_j^\mathsf{T}] = \mtx{0}$, and $\mathsf{E}[\vec{z}_i (\vec{z}_j^*)^\mathsf{T}] = \delta_{ij} \mtx{I}$).  Equation (\ref{eqn:entanglement}) corresponds to a macroscopic Bell state for $\ket{\Phi^+} = [\ket{HH}+\ket{VV}]/\sqrt{2}$ when $r$ is large.\cite{Iskhakov2011} The Jones vector for the vacuum mode entering the unused top port of \textsf{BS1} is
\begin{equation}
\vec{a}_3(t_0) = \sigma \vec{z}_3 \; .
\end{equation}
The random vectors $\vec{z}_1$, $\vec{z}_2$, and $\vec{z}_3$ constitute the initial hidden variables of this classical model.  Four other random vectors, denoted $\vec{z}'_1, \ldots, \vec{z}'_4$, will be used to model vacuum modes arising from the beam blockers.

Let us now consider the transformation of these vectors through times $t_1$, $t_2$, and $t_3$.  For $\vec{a}_1$ there is no change, so $\vec{a}_1(t_3) = \vec{a}_1(t_0)$.  We shall therefore focus on vectors $\vec{a}_2$ and $\vec{a}_3$.  At time $t_1$, after passing through the components \textsf{BS1}, \textsf{PD1}, \textsf{M1}, \textsf{M2}, and (possibly) \textsf{BB1}, \textsf{BB2}, we have
\begin{align}
\vec{a}_2(t_1) &= b_2 \left[ \sqrt{R_1} \, \vec{a}_2(t_0) - \sqrt{T_1} \, \vec{a}_3(t_0) \right] \notag \\
&\quad + (1-b_2) \sigma \vec{z}'_2 \\
\vec{a}_3(t_1) &= b_1 \, e^{i\theta_1} \left[ \sqrt{T_1} \, \vec{a}_2(t_0) + \sqrt{R_1} \, \vec{a}_3(t_0) \right] \notag \\
&\quad + (1-b_1) \sigma \vec{z}'_1 \; .
\end{align}
At time $t_2$, following \textsf{BS2}, \textsf{PD2}, \textsf{M3}, \textsf{M4}, and (possibly) \textsf{BB3}, \textsf{BB4}, we have
\begin{align}
\vec{a}_2(t_2) &= b_3 \, e^{i\theta_2} \left[ \sqrt{R_2} \, \vec{a}_2(t_1) - \sqrt{T_2} \, \vec{a}_3(t_1) \right] \notag \\
&\quad + (1-b_3) \sigma \vec{z}'_3 \\
\vec{a}_3(t_2) &=b_4 \left[ \sqrt{T_2} \, \vec{a}_2(t_1) + \sqrt{R_2} \, \vec{a}_3(t_1) \right] \notag \\
&\quad + (1-b_4) \sigma \vec{z}'_4 \; .
\end{align}
Finally, after \textsf{BS3} we have
\begin{align}
\vec{a}_2(t_3) &= \sqrt{T_3} \, \vec{a}_2(t_2) + \sqrt{R_3} \, \vec{a}_3(t_2) \\
\vec{a}_3(t_3) &= \sqrt{R_3} \, \vec{a}_2(t_2) - \sqrt{T_3} \, \vec{a}_3(t_2) \; .
\end{align}

Measurements at the three detectors are modeled as threshold crossing events.\cite{LaCour&Williamson2020}  We define the event $D_1$ as the subset of hidden-variable states such that a detection occurs on detector \textsf{D1} (regardless of the other two detectors).  We define $D_2(\vec{b})$ and $D_3(\vec{b})$ similarly for detectors \textsf{D2} and \textsf{D3}, respectively, but note that these subsets may depend upon the measurement configuration, as specified by the vector $\vec{b}$.  More specifically, for a given detection threshold $\gamma \ge 0$, these events are defined as follows:
\begin{align}
D_1 &= \left\{ \vec{z}_1, \ldots, \vec{z}'_4 : \|\vec{a}_1(t_3) \| > \gamma \, \right\} \; , \\
D_2(\vec{b}) &= \left\{ \vec{z}_1, \ldots, \vec{z}'_4 : \|\vec{a}_2(t_3) \| > \gamma \, \right\} \; , \\
D_3(\vec{b}) &= \left\{ \vec{z}_1, \ldots, \vec{z}'_4 : \|\vec{a}_3(t_3) \| > \gamma \, \right\} \; .
\end{align}
Thus, $N_{t_1,t_2,t_3}(+,-,-)$, say, would be the number of outcomes observed for the coincident detection event $D_1 \cap \overline{D_2(1,0,0,1)} \cap D_3(1,0,0,1)$, where the overline indicates the complementary set.

As an aside, the quantum prediction, assuming an initial state of $\ket{\Psi(t_0)} = [\ket{H H \varnothing} + \ket{V V \varnothing}]/\sqrt{2}$, where $\varnothing$ represents the ``downward'' vacuum mode entering \textsf{BS1}, is given by the final (unnormalized) state
\begin{multline}
\ket{\Psi(t_3)} = \alpha_{+} \left[ \frac{\ket{H H \varnothing}+\ket{V V \varnothing}}{\sqrt{2}} \right] \\
 + \alpha_{-} \left[ \frac{\ket{H \varnothing H}+\ket{V \varnothing V}}{\sqrt{2}} \right] \; ,
\end{multline}
where
\begin{equation}
\begin{split}
\alpha_{+} &= b_2 b_3 \, e^{i\theta_2} \sqrt{R_1 R_2 T_3} - b_1 b_3 \, e^{i(\theta_1+\theta_2)} \sqrt{T_1 T_2 T_3} \\
&\quad + b_2 b_4 \sqrt{R_1 T_2 R_3} + b_1 b_4 \, e^{i\theta_1} \sqrt{T_1 R_2 R_3}
\end{split}
\end{equation}
and
\begin{equation}
\begin{split}
\alpha_{-} &= b_2 b_3 \, e^{i\theta_2} \sqrt{R_1 R_2 R_3} - b_1 b_3 \, e^{i(\theta_1+\theta_2)} \sqrt{T_1 T_2 R_3} \\
&\quad - b_2 b_4 \sqrt{R_1 T_2 T_3} - b_1 b_4 \, e^{i\theta_1} \sqrt{T_1 R_2 T_3} \; .
\end{split}
\end{equation}
In general, $|\alpha_{+}|^2 + |\alpha_{-}|^2 \neq 1$, reflecting the projective nature of the beam blockers.  For certain combinations of $b_1, \ldots, b_4$, however, these values will sum to unity.  This is true, in particular, for the combinations used in the experiment.  This fact provides a justification for normalizing the counts in each of the three types of experiments.  Thus, for example, we would expect that $P_{t_1,t_2,t_3}(+,-,-) \approx |\alpha_{-}|^2 = T_1 R_2 T_3$ for $\vec{b} = (1,0,0,1)$.


\section{Simulation Results}
\label{sec:results}

Simulations were performed using the classical model described in the previous section.  For each measurement configuration $\vec{b}$ a total of $N = 2^{20} \approx 10^6$ random realizations of $\vec{z}_1, \ldots, \vec{z}_3$ and $\vec{z}'_1, \ldots, \vec{z}'_4$ were used.  (The outcomes are similar whether the same or independent draws are used.)  For each configuration $\vec{b}$, we may define the set of coincident detection events
\begin{subequations}
\begin{align}
E_{+}(\vec{b}) &= D_1 \cap D_2(\vec{b}) \cap \overline{D_3(\vec{b})} \; , \\
E_{-}(\vec{b}) &= D_1 \cap \overline{D_2(\vec{b})} \cap D_3(\vec{b}) \; .
\end{align}
\end{subequations}
Let us denote the union of these two disjoint events by $E(\vec{b}) = E_{+}(\vec{b}) \cup E_{-}(\vec{b})$, which represents the set of valid measurement outcomes upon which to post-select for this measurement context.  Each random realization contained in $E_{\pm}(\vec{b})$ contributes to one or more of the counts $N_{t_1,t_3}(q_1,\pm)$, $N_{t_2,t_3}(q_2,\pm)$, or $N_{t_1,t_2,t_3}(q_1,q_2,\pm)$, depending upon the value of $\vec{b}$.  Thus, although $E_{+}(\vec{b})$ and $E_{-}(\vec{b})$ are mutually exclusive, $E_{+}(\vec{b})$ and $E_{-}(\vec{b}')$, for $\vec{b}' \neq \vec{b}$, need not be.

Considering $E_{\pm}(\vec{b})$ as finite sets and denoting by $\mu[ \, \cdot \,]$ the counting measure, we have
\begin{subequations}
\begin{align}
N_{t_1,t_3}(+,\pm) &= \mu[ E_{\pm}(1,0,1,1) ] \; , \\
N_{t_1,t_3}(-,\pm) &= \mu[ E_{\pm}(0,1,1,1) ] \; .
\end{align}
\end{subequations}
We determine $N_{t_2,t_3}(q_2,q_3)$ and $N_{t_1,t_2,t_3}(q_1,q_2,q_3)$ in a similar manner.  From these counts, the various PMFs and correlations may then be computed, as previously described in Sec.\ \ref{sec:experiment}.

For a maximal violation of the Leggett-Garg inequality, the values $\theta_1 = \theta_2 = 0$ and $T_1 = 0.5$, $T_2 = T_3 = 0.75$ were used.  Good agreement with the experimental results of Ref.\ \onlinecite{Joarder2022} was found for $r = 0.3$ and $\gamma = 2.0$, with a mean value of $\bar{K} = 1.37 \pm 0.02$ obtained.  (The entire experiment was repeated 30 times to estimate the mean and standard deviation.)  For the same parameter settings, a value of $\bar{W} = 0.044 \pm 0.004$ was obtained for a violation of the WLGI.

A violation of the Leggett-Garg inequality in a photonic experiment suggests low detection efficiency may be to blame.  We may define this efficiency as the ratio of coincident-to-heralding detections, but there are several different measurement contexts for which this may be defined.  If no beam blockers are used, then we may consider the set
\begin{equation}
\Lambda_{t_3} = E(1,1,1,1)
\end{equation}
of all realizations for which there is a coincident detection and define the detection efficiency to be
\begin{equation}
\eta_{t_3} = \frac{\mu[E(1,1,1,1)]}{\mu[D_1]} \; .
\end{equation}
Note that this is actually a conditional probability, since $D_1 \subseteq E(1,1,1,1)$.  Other measurement contexts can also be considered, but they do not always yield a total probability of one, even ideally, due to the beam blockers.  For the particular combinations we are considering, however, the ideal probabilities do add to one, so it makes sense to consider these alternatives.  In light of this, we may consider the sets
\begin{equation}
\Lambda_{t_1,t_3} = E(1,0,1,1) \cup E(0,1,1,1) \; ,
\end{equation}
\begin{equation}
\Lambda_{t_2,t_3} = E(1,1,1,0) \cup E(1,1,0,1) \; ,
\end{equation}
and
\begin{multline}
\Lambda_{t_1,t_2,t_3} = E(1,0,1,0) \cup E(1,0,0,1) \\ \cup E(0,1,1,0) \cup E(0,1,0,1)
\end{multline}
and define the detection efficiencies
\begin{equation}
\eta_{t_1,t_3} = \frac{\mu[\Lambda_{t_1,t_3}]}{\mu[D_1]} \le \sum_{q_1,q_3} \frac{N_{t_1,t_3}(q_1,q_3)}{N_1} \; ,
\end{equation}
\begin{equation}
\eta_{t_2,t_3} = \frac{\mu[\Lambda_{t_2,t_3}]}{\mu[D_1]} \le \sum_{q_2,q_3} \frac{N_{t_2,t_3}(q_2,q_3)}{N_1} \; ,
\end{equation}
and
\begin{equation}
\eta_{t_1,t_2,t_3} = \frac{\mu[\Lambda_{t_1,t_2,t_3}]}{\mu[D_1]} \le \sum_{q_1,q_2,q_3} \frac{N_{t_1,t_2,t_3}(q_1,q_2,q_3)}{N_1} \; .
\end{equation}
The upper bounds represent experimentally accessible values; however, with this classical model it is possible to take the union over counterfactually distinct measurement contexts and compute the efficiencies directly.

For the aforementioned parameter settings, an overall detection efficiency (i.e., without any beam blockers) of $\eta_{t_3} = 0.0809 \pm 0.0005$ was found.  By contrast, with one of the first two (last two) beam blockers inserted the efficiency dropped to $\eta_{t_1,t_3} = 0.0530 \pm 0.0005$ ($\eta_{t_2,t_3} = 0.0657 \pm 0.004$).  With two beam blockers in place, the value $\eta_{t_1,t_2,t_3} = 0.0587 \pm 0.0003$ was obtained.  All four detection efficiencies are statistically different from one another.  It was also found, perhaps not surprisingly, that the sets $\Lambda_{t_1,t_3}$, $\Lambda_{t_2,t_3}$, and $\Lambda_{t_1,t_2,t_3}$ are all distinct.  In other words, hidden variables that produce a coincidence in one group of measurement contexts might not do so in another context.

The corresponding double-detection probabilities, defined by
\begin{equation}
\delta(\vec{b}) = \frac{\mu[ D_1 \cap D_2(\vec{b}) \cap D_3(\vec{b}) ]}{\mu[D_1]} \; ,
\end{equation}
were also measured and found to be about two orders of magnitude smaller than the detection efficiencies.  For example, $\delta(1,1,1,1) = (3.8 \pm 0.4) \times 10^{-4}$ and $\delta(1,1,0,1) = (4.9 \pm 0.4) \times 10^{-4}$ are the largest such values.  Double-detection events are commonly interpreted as indicating the presence of two or more photons in the interferometer.  Such situations can given rise to the so-called multi-photon emission loophole, under which a macroscopic realist model could show a violation of the Leggett-Garg inequality.\cite{Joarder2022}  In the extreme case, classical (i.e., bright) coherent light can show such violations.\cite{Zhang2021}  In the present case, double-detection events are extremely rare compared to single-detection events and, hence, are insufficient to produce the large violation observed.

\begin{figure}[ht]
\includegraphics[width=\columnwidth]{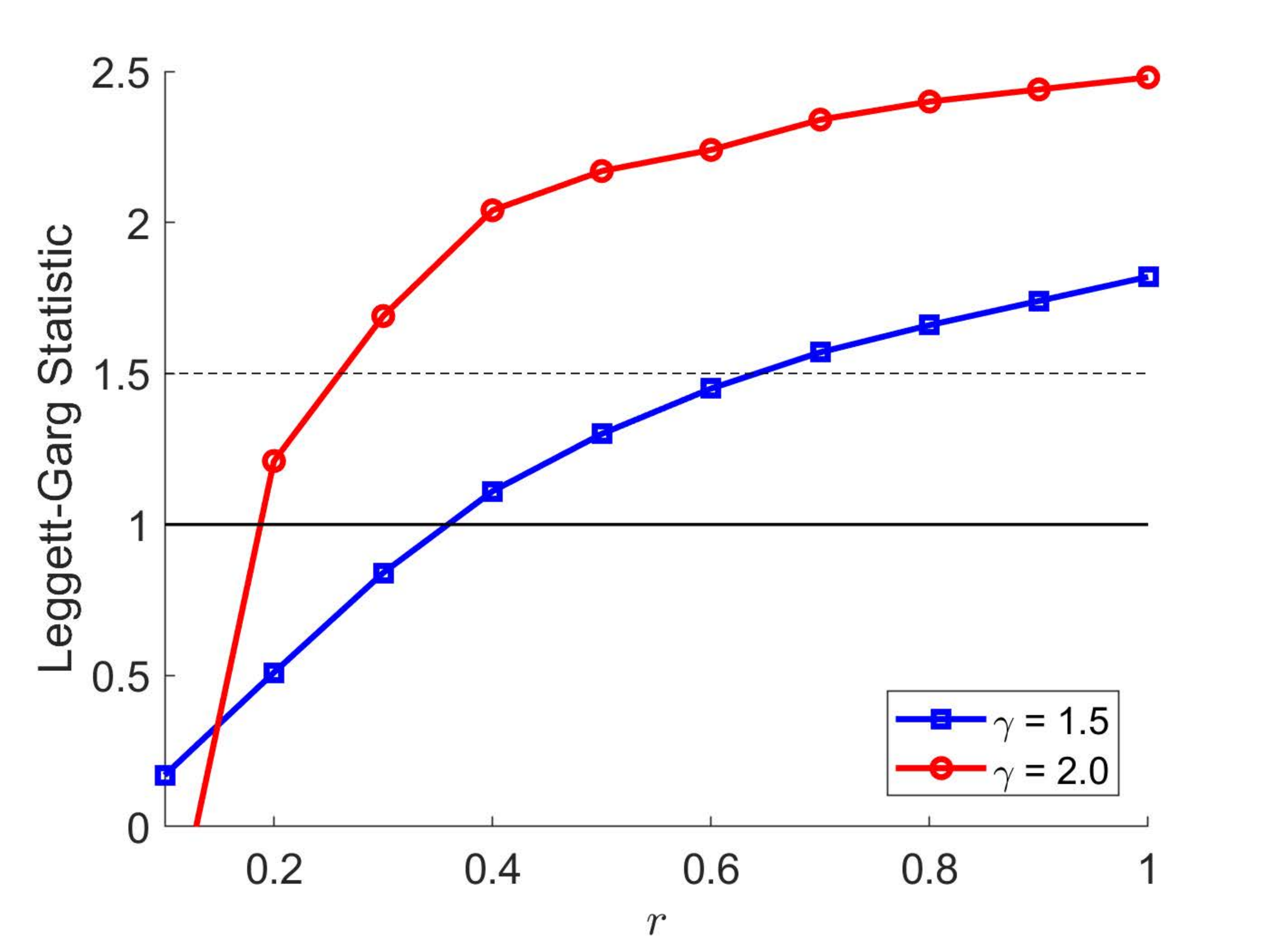}
\caption{(Color online) Plot of the Leggett-Garg statistic, $K$, versus the squeezing strength, $r$, of the entanglement source.  The blue squares and red circles correspond to detection thresholds of $\gamma = 1.5$ and $\gamma = 2.0$, respectively.  The solid horizontal black line is the upper bound for macroscopic realism, while the dash horizontal black line is the maximum value predicted by quantum mechanics.}
\label{fig:LGI}
\end{figure}

Having established these initial observations for $r = 0.3$ and $\gamma = 2.0$, an examination was made of the dependence of the $K$ statistic on these two parameters, taking $0 \le r \le 1$ and $\gamma \in \{ 1.5, 2.0 \}$.  Figure \ref{fig:LGI} illustrates the findings.  There we see that, generally speaking, larger values of $r$ and $\gamma$ tend to produce larger violations.  Indeed, for sufficiently large values a violation of the quantum mechanical upper bound of 1.5 is even found to be possible.  This, of course, is due to post-selection on coincidences in extreme parameter regimes, which is known to be capable of producing such anomalous effects.\cite{Chu2016}

\section{Detection Efficiency Loophole}
\label{sec:loophole}

The experimental procedure and analysis used in this work is similar to that of Ref.\ \onlinecite{Joarder2022}, where the authors claim that their use of beam blockers and marginal PMFs renders the detection loophole irrelevant for any nonzero efficiency.  ``Thus, the violations of both LGI and WLGI measured in this way cannot be reproduced by the hidden variable model, whatever the detector efficiency, thereby rendering \emph{the detection efficiency loophole irrelevant in this context.} [Emphasis in original]''\cite{Joarder2022}  It is important to understand that what they describe is a \emph{macrorealistic} hidden-variable model and, as such, their arguments say nothing of the possibility for a local realist interpretation.  Although this restriction is reasonable for the purpose of rejecting a macrorealist interpretation, it is insufficient for rejecting a local realist interpretation of the experiment.  For the latter, the detection efficiency loophole is essential.

In Ref.\ \onlinecite{Joarder2022}, it is supposed that the hidden-variable model is such that one may identify a subset $\Lambda_3$ of hidden-variable states  for which a photon is detected at time $t_3$.  However, there are several measurement contexts under which this may occur.  In particular, four sets, $\Lambda_{t_3}$, $\Lambda_{t_1,t_3}$, $\Lambda_{t_2,t_3}$, and $\Lambda_{t_1,t_2,t_3}$, were identified for which a valid, heralded measurement at $t_3$ is made.  Under the assumption of macrorealism, all four sets are identical; however, the numerical investigations of Sec.\ \ref{sec:results} have shown that they may be distinct.

In a similar manner, the authors of Ref.\ \onlinecite{Joarder2022} define a single detector efficiency $\eta$ for their hidden-variable model.  If we let $\mu[ \,\cdot\, ]$ denote the probability measure for the hidden-variables, then this efficiency would be defined as $\eta = \mu[ \Lambda_3 | D_1 ]$.  (Previously, $\mu$ was defined as the counting measure; here it is simply generalized.)  As previously noted the detection efficiency depends upon the measurement context, so this is only valid under an assumption of macrorealism.  Three conceptually distinct detection efficiencies may be defined that, in the current context, may be written
\begin{equation}
\eta_{t_i,t_3} = \mu[ \Lambda_{t_i,t_3} | D_1 ] \; ,
\end{equation}
for $i \in \{1,2\}$, and
\begin{equation}
\eta_{t_1,t_2,t_3} = \mu[ \Lambda_{t_1,t_2,t_3} | D_1 ] \; .
\end{equation}
As noted previously, it was indeed found that these detection efficiencies can be meaningfully different from one another.  Thus, the assumption of a single efficiency for all measurement contexts is likewise unwarranted for a local realist, but not macrorealist, model.

To see the import of these assumptions, it is perhaps simplest to consider the WLGI.  (A similar argument can be made for the LGI.)  Recall that
\begin{equation}
W = P_{t_1,t_3}(-,+) -  P_{t_2,t_3}(-,+) - P_{t_1,t_2}(-,+) \; ,
\end{equation}
where, now,
\begin{equation}
\begin{split}
P_{t_1,t_3}(-,+) &= \mu[ E_+(0,1,1,1) | \Lambda_{t_1,t_3}] \\
&= \mu[ E_+(0,1,1,1) | D_1 ] / \eta_{t_1,t_3} \; ,
\end{split}
\end{equation}
\begin{equation}
\begin{split}
P_{t_2,t_3}(-,+) &= \mu[ E_+(1,1,0,1) | \Lambda_{t_2,t_3}] \\
&= \mu[ E_+(1,1,0,1) | D_1 ] / \eta_{t_2,t_3} \; ,
\end{split}
\end{equation}
and
\begin{equation}
\begin{split}
P_{t_1,t_2}(-,+) &= \sum_{\pm} \mu[ E_{\pm}(0,1,1,0) | \Lambda_{t_1,t_2,t_3} ] \\
&= \sum_{\pm} \mu[ E_{\pm}(0,1,1,0) | D_1 ] / \eta_{t_1,t_2,t_3} \; .
\end{split}
\end{equation}
Now, unlike the true WLGI, $W$ does not have an upper bound of zero since $P_{t_1,t_3}$ and $P_{t_2,t_3}$, unlike $P_{t_1,t_2}$ are not marginal PMFs derived from $P_{t_1,t_2,t_3}$.

By contrast, if we use the marginal probabilities
\begin{multline}
\tilde{P}_{t_1,t_3}(-,+) \\
= \frac{\mu[ E_+(0,1,1,0) | D_1 ] + \mu[ E_+(0,1,0,1) | D_1 ]}{\eta_{t_1,t_2,t_3}}
\end{multline}
and
\begin{multline}
\tilde{P}_{t_2,t_3}(-,+) \\
= \frac{\mu[ E_+(1,0,0,1) | D_1 ] + \mu[ E_+(0,1,0,1) | D_1 ]}{\eta_{t_1,t_2,t_3}} \; ,
\end{multline}
then the WLGI of Eqn.\ (\ref{eqn:WLGI}) follows quite readily.

The inequality results from the relationship between the three probabilities arising from the common joint PMF from which they are derived and the common efficiency in all terms.  Thus, whether $W$ is bounded by zero or not depends only on whether marginal probabilities are used, not on the values of the detection efficiencies.  Of course, experimentally, $W$ is computed as it is defined, using directly measured probabilities rather than marginally inferred ones.  It is only if one restricts to hidden-variable models obeying macrorealism that the two become identical.  Similar comments apply to $K$ and the LGI.


\section{Conclusion}
\label{sec:conclusion}

This paper considered tests of macrorealism in the context of a quantum optical interferometry experiment using heralded photon detection and negative measurements.  It may be concluded that, while the assumption of macrorealism is inconsistent with such experiments, they do not demonstrate uniquely quantum behavior.  This was shown explicitly by obtaining an LGI violation using a classical model of entangled light and a deterministic model of single-photon detectors.  Such a result was possible only by following the same experimental measurement and analysis procedures.  In particular, post-selection by the use of heralded detections was found to introduce contextuality through a violation of the fair sampling hypothesis.  This, in turn, allowed for agreement with quantum predictions, for which contextuality is needed.

Although the model discussed in this paper is intended to describe quantum light, it would be straightforward to implement a classical analogue of the experiment using a pair of bright laser light sources.  Intensity and polarization modulators could be used to reproduce the correlated stochastic behavior described in the model, and a power meter with a discriminator could be used to represent a single-photon detector.


\begin{acknowledgments}

The author would like to thank Urbasi Sinha for a helpful description of her group's experimental analysis.  He would also like to thank Denys Bondar and Wenlei Zhang for an enlightening discussion of their experiments with classical light.  This work was supported in part by the Office of Naval Research under Grant No.\ N00014-18-1-2107.

\end{acknowledgments}


\section*{Author Declarations}

\subsection*{Conflict of Interest}

The author declares no conflict of interest.


\section*{Data Availability}

The data that support the findings of this study are available from the corresponding author upon reasonable request.





%


\end{document}